\documentstyle[11pt]{article}

\def\sect{\section}
\def\EQ{\begin{equation}}
\def\EN{\end{equation}}
\def\bea{\begin{eqnarray}}
\def\ena{\end{eqnarray}}
\newcommand{\vs}[1]{\vspace{#1 mm}}
\def\z{{\bar z}}
\def\nn{\nonumber \\}
\def\pa{\partial}
\def\a{\alpha}
\def\b{\beta}
\def\c{\gamma}
\def\d{\delta}
\def\e{\epsilon}
\def\p{\varphi}
\def\s{\sigma}
\def\L{\Lambda}
\def\ss{s_\L}
\def\bel{\mu^z_\z}
\def\cz{c^z}

\def\pz{\pa_z}
\def\pzb{\pa_\z}
\def\w{w(m)}
\def\mz{\mu^z_\z}
\def\F{f^a_{bc}}
\def\G{g^a_{bc}}
\def\Mz{M_{\z}}
\def\mh{\hat {\mu}^z}
\def\Mh{\hat M}
\def\Mt{\tilde M}
\def\gg{\c^\shalf}
\def\gr{\a^{\shalf}_\z}
\renewcommand{\t}{\theta}
\newcommand{\shalf}{\frac{1}{2}}
\begin{document}
\begin{titlepage}
\begin{center}

\hfill  PAR--LPTHE 96--37, OU-HET 248 \\
\hfill  hep-th/9609207 \\
        [.5in]

{\large\bf
WORLDSHEETS WITH EXTENDED SUPERSYMMETRY
}\\[.5in]

{\bf Laurent Baulieu}\footnote{email address:
baulieu@lpthe.jussieu.fr} \\
{\it LPTHE, Universit\'es Paris VI - Paris VII, Paris, France}\footnote{
Universit\'es Paris VI - Paris VII, URA 280 CNRS,
4 place Jussieu, F-75252 Paris CEDEX 05, FRANCE.}\\
and \\
{\it Yukawa Institute for Theoretical Physics,
 Kyoto University, Kyoto 606, Japan} \\

\vs{5}
{\bf Nobuyoshi Ohta}\footnote{email address:
ohta@phys.wani.osaka-u.ac.jp} \\
{\it Department of Physics, Osaka University, Toyonaka, Osaka 560, Japan}

\end{center}

\vs{5}
\centerline{{\bf{Abstract}}}
\vs{5}

We determine the equations which govern the gauge symmetries of worldsheets
with local supersymmetry of arbitrary rank $(N,N')$, and their possible
anomalies. Both classical and ghost conformally invariant multiplets of
the left or right sector are assembled into the components of a single
$O(N)$-superfield. The component with ghost number zero of this superfield
is the $N$-supersymmetric generalization of the Beltrami differential.
In a Lagrangian approach, and after gauge-fixing, it becomes the super-moduli
of Riemann surfaces coupled to local supersymmetry of rank $N$. It is also
the source of all linear superconformal currents derived from ordinary
operator product techniques. The interconnection between BRST invariant
actions with different values of $N\geq 3$, and  their possible link to
topological 2D-gravity coupled to topological sigma models,  are shown by
straightforward algebraic considerations.  
\end{titlepage}

\newpage
\renewcommand{\thefootnote}{\arabic{footnote}}
\setcounter{footnote}{0}

\sect{Introduction}

The quantum theories based on the symmetries of 2D-gravity coupled to local
supersymmetry have proven to be extremely powerful to describe physical
phenomena.

The chosen rank of the supersymmetry varies from one application to the other.
However, very intriguing relations have been shown to exist between the
various theories with different values of $N$. In particular, the existence of
a sort of embedding of $N$-theories into $(N+1)$-theories has been shown from
several points of view~\cite{BEV,BOP,BGR}. This seems to privilege the
theory with $N=\infty$, while the $N=3,4$ theories play a boundary role
between the theories with a relativistic matter interpretation, that is
with $N=0,1,2$ and the others.

Here we will show that a kind of universality also exists for the mathematical
description of worldsheets with extended supersymmetry, with a surprisingly
simple algebraic structure. We point out the possibility of generalizing
the very old notion of Beltrami parametrization of conformally invariant
Riemann surfaces to the case where reparametrization invariance is coupled
to local worldsheet supersymmetry of arbitrary rank $(N,N')$. We find that
the full multiplet of conformally invariant 2D-supergravity of rank $N$
in the left sector is made of all possible 2D gauge fields valued in
the antisymmetric tensor representations of $O(N)$, with similar properties
in the right sector. Moreover the complete BRST equations of this theory,
and thus the algebra of its gauge transformations, follow from a universal
vanishing curvature condition in superspace. These will be described
in sect.~2.

This formulation of the gauge symmetries 
of 2D-supergravities as a vanishing curvature condition enables us to
identify their ghost actions as a $BF$ systems, which needless to say,
emphasizes the topological aspects of these theories. We will briefly
elaborate on this by using the Batalin-Vilkovisky (BV) framework~\cite{BV}
in sect.~3.

The BRST symmetry of the ghosts that we obtain coincides with that one obtains
from the usual operator product expansion (OPE) treatments of linear
superconformal algebra (SCA). Since we  define the complete set of gauge
fields of conformal 2D-supergravity, we can use them as background gauge
fields. This gives quite a simple and clear reversed construction of the
currents of SCA by use of these background gauge fields as the
sources of the currents.

When gauge fixing the symmetries of a worldsheet with $(N,N')$ local
supersymmetry, the gauge fields of 2D-supergravity should not be put equal
to zero, since this would result in an over gauge-fixing for higher genera.
Rather, they can be at most set equal to constant backgrounds over which
one should integrate, by taking also into account the remaining modular
invariance. Our presentation has thus also the advantage of defining all
(super)-moduli of extended 2D-supergravities in a well-defined framework.

We illustrate some of our results in the cases of $N=2,3$ and (large) $N=4$
supersymmetry.

We give a combinatoric proof that all pure 2D-supergravities with $N\geq 3$
are anomaly free, which would only allow their couplings to topological matter,
in contrast with the cases $N=2,1,0$ for which critical matter must be
introduced to compensate for the anomaly. We also elaborate on the question
of the consistent anomaly, for which we give explicit expressions for $N=2,3$,
which generalize the known results for $N=0,1$.

Finally, we demonstrate the embedding properties between $N-$ and
$(N+1)-$theories, by showing that for $N\geq 3$ the BRST invariant action of
the former can be considered as part of the latter. We also suggest a link 
between these theories and topological 2D-gravity coupled to topological
sigma models with bosonic and fermionic coordinates.

\sect{Conformally invariant parametrization for 2D-supergravities}

The Beltrami parametrization consists in expressing the squared length
elements of the worldsheet as
\bea
d\s^2=\exp \p (dz+\mz d \z)(d\z+\mu^\z_ z d z).
\ena
This parametrization is very natural because the Beltrami differential
$\mu^\z_z$ undergoes the reparametrization symmetry under a factorized form,
and is dilatation invariant. More precisely, the BRST symmetry transformation
of $\mz(z,\z)$ only involves a ghost $\cz(z,\z)$ (which can be identified as
a suitable combination of the two components of the ordinary reparametrization
ghost field). In ref.~\cite{BB}, the extension of the Beltrami parametrization
to the case of $N=1$ 2D-supergravity was found, by introducing the conformally
invariant part of the gravitino $\gr(z,\z)$, with ghost $\gg(z,\z)$, and
the following factorized BRST algebra
\bea
s\bel&=&\pzb\cz+\cz\pz\bel-\bel\pz\cz - \shalf \gr\gg,\nn
s\gr&=&-\pzb\gg-\shalf \gg\pz\bel+ \bel \pz\gg +\cz\pz\gr+\shalf\gr\pz\cz,\nn
s\cz&=&\cz\pz\cz-{1 \over 4} \gg\gg,\nn
s\gg&=&\cz\pz\gg - \shalf\gg\pz\cz,
\label{symmtwo}
\ena

This algebra has remarkable algebraic properties, and we will show how it can
be generalized to the case of extended supersymmetry of any given rank $N$.
Eq.~(\ref{symmtwo}) has a simple superfield formulation.
Indeed, by introducing one single Grassmann variable $\t$, these BRST
equations can be written as
\EQ
\label{unif}
\hat d \Mh^z-\Mh^z\pz\Mh^z+{1 \over 4}(D_\t \Mh^z)^2=0,
\EN
where $D_\t =\pa_\t + \t\pz$ and the BRST transformation operator $s$ and the
differential $d=dz\pz+d\z\pzb$ are unified into $\hat d=d+s$ while the
holomorphic part of the conformally invariant classical gauge fields of
$N=1$ 2D-supergravity and of their ghosts are assembled into the superfield
$\Mh(z,\z,\theta)$ as $\Mh^z= M^z+C^z$ with
\EQ
M^z(z,\z,\theta)= dz+\bel(z,\z)d\z +\t \gr (z,\z)d\z , \quad \quad 
C^z(z,\z,\theta)= \cz(z,\z) +\t \gg(z,\z).
\EN
By expanding (\ref{unif}) in ghost number and powers of $\t$ one recovers
(\ref{symmtwo}). Notice that the property $\hat d^2 =s^2=0$ implies
the $N=1$ supersymmetry relation $D_\t^2 =\pz.$

Eq.~(\ref{unif}) can be understood as a direct consequence of the vanishing
of the torsion in superspace, without any reference to the OPE techniques.
Since the Beltrami differential and the gravitino are
the source of the energy-momentum tensor and of the supersymmetry current,
this can be seen as the possibility for a reversed, and perhaps more
geometrical, construction of superconformal quantum field theory.

We now generalize this to higher rank supersymmetry. We use the natural
framework for extended 2D-supergravity, which,
for the holomorphic sector, is the $N$-superspace with
coordinates $(z, \t^i)$, $i=1,\cdots, N$~\cite{ademollo,schou}.

We define the generalization in $N$-superspace of the one-form
which unifies the Beltrami differential and its ghost as
\bea
\Mh^z(z,\z,\theta) &=& dz+\bel d\z+\cz + \t^i\left (\a^{\shalf}_{\z i} d\z
 +\gg_i \right)
 + \shalf\ \sum_{ij}\ \t^i\ \t^j \left( C_{\z ij}^0 d\z +c_{ij}^0\right) \nn
&& +\sum_{p=3}^N\ 
\sum_{1\leq i_1\dots i_p\leq N }\ \ {1\over p!}\ 
\t^{i_1}\cdots\t^{i_p} \left ( C_{\z i_1\dots i_p}^{1-{p\over 2}}d\z+
 c_{i_1\dots i_p}^{1-{p\over 2} }\right).
\ena

The Beltrami differential $\bel(z,\z)$ and the classical fields
$C_{\z i_1\dots i_p}^{1-{p\over 2}}(z,\z)$, $\ p=1,\cdots, N$, (with ghost
number zero) define the conformally invariant left sector of the $(N,N')$
2D-supergravity. The anticommuting fields $\a^{\shalf}_{\z i}(z,\z)$ are
identified as the holomorphic parts of the $N$ gravitini (eventually they
will be the sources of the $N$-supersymmetry currents) and the commuting
fields $ C_{\z ij}^0(z,\z)$ are the components along $\z$ of the commuting
$O(N)$ gauge field which gauges the $O(N)$ rotations. The other classical
fields contained in $\Mh^z$, $C_{\z i_1\dots i_p}^{1-{p\over 2}}(z,\z)$,
$\ p=2,\cdots, N$, gauge the internal fermionic and bosonic symmetries of
2D-supergravity of rank $N$. Thus, the holomorphic components of the
2D-supergravity multiplet are quite simply identified as the antisymmetric
tensor representations of $O(N)$. The fields
$c_{i_1\dots i_p}^{1-{p\over 2}}(z,\z)$ are their ghosts, with conformal
weight ${p\over 2} -1$. $C_{\z i_1\dots i_p}^{1-{p\over 2}}(z,\z)$ and
$c_{i_1\dots i_p}^{1-{p\over 2}}(z,\z)$ have opposite statistics.

It is easy to verify that, as required by supersymmetry, the number of
commuting fields in this multiplet equals the number of anticommuting fields.
Indeed, the number of independent fields contained in the $\t$ polynomials
$\t^{i_1}\cdots\t^{i_p} C_{\z i_1\dots i_p}^{1-{p\over 2}}$ of rank $p$
is equal to $_N C_p$. Since $ 0=(1-1)^N$, one has $\sum_p{}_NC_{2p}
=\sum_p{}_N C_{2p+1}$, which shows the required equality between the number
of bosons and fermions, necessary to ensure the eventual closure of the
2D-supergravity algebra.

The BRST symmetry is defined by the following straightforward generalization
of eq.~(\ref{unif}):
\bea
\label{unifN}
\hat d \Mh^z &=&
\Mh^z\pz\Mh^z-{1 \over 4}\sum_{i=1}^N (D_i \Mh^z)^2, \\
\Mh^z &=& M^z_\z d\z + C^z.
\ena
One has the closure relation $\hat d^2=0$, that is $s^2=0$, if and only if
\EQ
D_i D_j+ D_j D_i=2\d_{ij}\pz.
\label{symN}
\EN
Therefore, one has $D_i =\pa_{\t^i} + \t^i\pz$ in eq.~(\ref{unifN}).

The vanishing curvature condition (\ref{unifN}) can also be
understood as a realization of the abstract algebra (\ref{symN}).
The BRST transformations of the classical and ghost superfields
are then easily extracted from (\ref{unifN}) as
\bea
\label{sM}
sM^z_\z &=& \pzb C^z+C^z\pz M^z_\z-M^z_\z\pz C^z
 - \shalf \sum_{i=1}^N D_i C^z D_i M^z_\z,\\
s C^z &=& C^z \pz C^z - \frac{1}{4} \sum_{i=1}^N (D_i C^z)^2.
\label{sutran}
\ena
Eq.~(\ref{sM}) gives the classical gauge transformations of all components
of the 2D-supergravity multiplet, simply by changing the ghosts into
infinitesimal parameters with the opposite statistics.

Let us notice that one can deduce the full BRST algebra from
the sole knowledge of the BRST transformation of the
superfield $C^z$. Indeed, the complete BRST equations (\ref{unifN}), can be
directly obtained from (\ref{sutran}) with the substitutions
\EQ
\label{sub}
s \to d+s, \qquad
C^z \to M^z+C^z.
\EN
It follows that the determination of the ghost transformation obtained from
the OPE techniques in superconformal quantum field theory, as shown in
ref.~\cite{BOP}, would also indirectly permit the determination of the gauge
fields associated to these ghosts and of their transformation laws.

\sect{The antighosts and the gauge-fixed action}

Usually antighosts are directly introduced as conjugate variables to
the ghosts. This is a consistent approach since one is mainly interested
in superstring theory expressed in the (super)conformal gauge.
In order to remain in a geometrical framework, it is however interesting
to introduce in a gauge-independent way the conjugates of all fields
contained in $\Mh^z(z,\z,\theta)$. To do so, we will use the BV formalism~\cite{BV},
where the (super)antifields are naturally the duals to the
(super)fields~\cite{BBB}. The usual antighosts will be introduced afterwards
via an appropriate choice of the gauge function.

Let us denote the antifields of $M^z_\z$ and $C^z$ as $^* M_{zz} $ (with
ghost number $-1$) and $^*C_{zz\z}$ (with ghost number $-2$), respectively,
and define
\bea
^*\Mh_z(z,\z,\theta) = \ ^* \hspace{-.2mm} M_{zz}dz+\ ^* \hspace{-.2mm}
 C_{zz\z} dzd\z.
\ena
The $O(N)$ superspace decomposition of $^*\Mh_z$ is
\bea
^*\Mh_z(z,\z,\theta) &=& \frac{1}{N!}\e_{i_1\cdots i_N}\t^{i_1}\cdots\t^{i_N}
 \left( ^*\mu_{zz} d z+\ ^* \hspace{-.2mm} c_{zz\z} dzd\z \right) \nn
&& +\sum_{p=0}^{N-1}\ \frac{1}{p!}\e_{i_1\cdots i_N} \t^{i_1}\cdots\t^{i_p}\ 
 \left( ^*C_{zz i_{p+1}\dots i_N}d z
+ \ ^* \hspace{-.2mm}c_{zz\z{i_{p+1}\dots i_N}} dzd\z \right).
\ena

The invariant BV action which determines the BRST symmetry (\ref{unifN})
is the part with ghost number $g=0$ of
\bea
\label{bv}
I_{BV}=\int d^2 z d^N\t \ ^*\Mh_z \hat G^z,
\ena
where
\EQ
\hat G^z = d \Mh^z - \Mh^z\pz\Mh^z+{1 \over 4}\sum_{i=1}^N (D_i \Mh^z)^2.
\EN
This BV action satisfies a (first rank) master BV equation. This means
that it defines a nilpotent differential operator BRST, given by 
\EQ
s \Mh^z={{\d I_{BV}^{g=0}}\over{\d ^* \Mh_z}},\qquad
s\ ^* \hspace{-.2mm} \Mh_z={{\d I_{BV}^{g=0} }\over{\d \Mh^z}}.
\EN
The first equation is identical to the BRST transformation law found earlier
for $ \Mh^z$; the second equation, which expresses the BRST transformation
of antifields, implies eventually that the currents, in the superconformal
gauge, are BRST-exact.

This form of the action (\ref{bv}) (prior to any kind of gauge fixing)
indicates the rather deep connection of the theory with a topological BF type
system. Notice that it contains no classical part, since the part with ghost
number zero of $\hat G^z$ in $I_{BV}$ consists of ghosts only.

One naturally considers the gauge where one imposes that the gauge fields
contained in the expansion of $\Mh^z(z,\z,\theta)$ are set equal to a
background value, which we shall denote as $M^z_{\z,bg}(z,\z,\theta)$.
(The ``superconformal gauge"
is obtained for $M^z_{\z,bg}=0$.) This choice of gauge implies the
introduction of antighosts, with the following superfield expansion
\EQ
B_z(z,\z,\theta)= \frac{1}{N!}{{\e_{i_1\cdots i_N}} }\t^{i_1}\cdots\t^{i_N}
 b_{zz} dz + \frac{1}{(N-1)!}{{\e_{i_1 i_2\cdots i_N}} }\t^{i_2}\cdots\t^{i_N}
 \b_{zz}^{ i_1} dz +\cdots ,
\EN
and the following BV gauge function:
\EQ
Z_{GF}=\int d^2zd^N \t\ B_z(M^z_\z -M^z_{\z,bg}).
\EN
It implies
\EQ
M^z_\z = M^z_{\z,bg},\qquad
^* M_{zz}= {{\d Z_{GF}}\over{\d M^z_\z}} =B_z,\qquad
^*C_{zz\z}= {{\d Z_{GF}}\over{\d C^z}}=0.
\label{bg}
\EN

The BV action then reduces to an action $I_{GF}$ which only depends on the
ghosts and antighosts
\bea
I_{GF} &=& \int d^2z d^N\t\ B_z s M^z_\z |_{M^z_\z =M^z_{\z,bg}}\nn
&=& \int d^2z d^N\t\ B_z \left(\pzb C^z+C^z\pz M^z_{\z,bg}
 - M^z_{\z,bg}\pz C^z- \shalf\sum_{i=1}^N D_i C^z D_i M^z_{\z,bg}\right).
\label{action}
\ena
Let us give for completeness the expression of $I_{GF}$ after
integration upon the supercoordinates $\theta^i$:
\EQ
I_{GF} =\int d^2 z\left( b_{zz}\ s\bel
 + \sum_{i_1\cdots i_p}\beta_{zz}^{i_1\cdots i_p}\ s M^z_{\z\ i_1\cdots i_p}
 \right)_{M^z_\z =M^z_{\z,bg}}.
\label{act1}
\EN
The detailed expressions of $s \bel= \pa_\z c^z+\ldots$ and
$sM^z_{\z\ i_1\cdots i_p}= \pa_\z m^z_{\ i_1\cdots i_p}+\ldots$,
$1\leq p \leq N$, follow from the decomposition of eq.~(\ref{sM}). We will
shortly comment on the structure of these BRST transformations of the
fields $M^z_{\z\ i_1\cdots i_p}$ in component formalism.

\sect{The currents and the background symmetry}

We can define the following supercurrent from (\ref{action}):
\bea
J_z &=& {{\d I_{GF}\over{\d M^z_{\z,bg}}}}\nn
&=& \left(\frac{N}{2}-2 \right) B_z\pz C^z - \pz B_z C^z
 + \frac{(-)^{\e_B}}{2}\sum_{i=1}^N D_i B_z D_i C^z,
\label{sc}
\ena
where $\e_B$ is odd (even) integer for anticommuting (commuting) $B_z$.

The property that the above superfield currents (\ref{sc}) truly represent the
linear SCA, and that the currents that are obtained by its decompositions onto
the various $O(N)$ representations are (classically) conserved, is warranted
by the existence of the background gauge symmetry for
$I_{GF}$, which generalizes the background reparametrization invariance
of the purely bosonic case. Let us now determine this symmetry, using
the fact that we have at our disposal the background superfield $M^z_{GF}$.

It is most convenient to represent the infinitesimal $O(N)$ superfield
parameter of this background gauge symmetry by an $O(N)$ superfield ghost of
opposite statistics, denoted as $\L^z(z,\z,\theta)$. Then one can {\it
 ghostify} the infinitesimal background gauge transformations under the form
of an associated nilpotent generator $\ss$ defined by the following equation:
\bea
(d+s+\ss)\Mt^z &=& \Mt^z\pz\Mt^z
-{1 \over 4}\sum_{i=1}^M(D_i \Mt^z)^2, \nn
\Mt^z &=& \Mh^z + \L^z,
\ena
where one assigns a new distinct ghost numbers to $\ss$ and $\L$ with
$s\L^z=0$. The differential operators $s$ and $\ss$ anticommute. One can
find the transformation of the antighost $B$ under $\ss$ such that the
action $I_{GF}$ is invariant under the the background gauge symmetry,
$\ss I_{GF}=0$:
\bea
\ss M^z_{\z,bg} &=& \pzb \L^z+\L^z\pz M^z_{\z,bg}- M^z_{\z,bg}\pz\L^z
 - \shalf \sum_{i=1}^N D_i\L^z D_iM^z_{\z,bg}, \nn
\ss C^z &=& C^z\pz \L^z+\L^z\pz C^z- \shalf\sum_{i=1}^N D_i\L^z
D_i C^z, \nn
\ss B_z &=& \L^z\pz B_z + \left(2-\frac{N}{2}\right) \pz \L^z B_z
 - \shalf\sum_{i=1}^N D_i \L^z D_i B_z.
\ena
This background gauge symmetry determines the Slavnov identity of the
theory, and also ensures that the gauge-fixed action has the relevant
superconformal symmetry of rank $N$.

\sect{Pure component formalism and the cases $N=2,3$ and $4$}

We now show that the sole knowledge of a SCA in component formalism would
also provide the full information about the holomorphic Beltrami-like
decomposition of 2D-supergravity and the BRST quantization of worldsheets
with extended supersymmetry.

In full generality, a linear SCA is a super Lie algebra, whose structure
coefficient determines a BRST symmetry of the following form
\bea
\label{Q}
s \cz &=& \cz\pz\cz - \frac{1}{4} \c^i\c_i, \nn
sm^a&=&\cz\pz m^a-{\w} m^a\pz\cz+\F m^b m^c+\G m^b\pz m^c,
\ena
where $\F$ and $\G$ are (constant) structure coefficients such that $s^2=0$
on all fields and $\w$ is the conformal weight of the ghost $m^a$.
$s\c^i$ has an expression analogous to $s m^a$. It is convenient to
separate the reparametrization ghost $\cz$ and the local worldsheet
supersymmetry ghosts $\c^i$, $1\leq i\leq N$, from the rest of the ghosts
denoted as $m^a$ which are either odd or even and correspond to the
internal gauge bosonic and fermionic symmetries of the SCA.

The holomorphic components of the classical gauge fields can now be defined
directly. They are the Beltrami differential $\mz$, in correspondence
with $\cz$, the $N$ gravitini $\a^i_\z$, in correspondence with $\c^i$,
and the components of 1-forms along $d\z$, $\Mz ^ad\z$, in correspondence
with the ghosts $m^a$. One defines the unified objects
\bea
\cz \to \mh&=&dz+\mz d\z+\cz, \nn
\c^i \to \hat\a^i&=&\a ^i_\z d\z+\c^i,\nn
m^a \to \Mh^a&=& \Mz ^a d\z+m^a.
\label{unified}
\ena
Then the full BRST algebra is simply read off from (\ref{Q}) as
(for convenience we now incorporate the fields $\hat\a^i$ as a part of
the $\Mh^a$)
\bea
(d+ s) \mh &=& \mh\pz\mh - \frac{1}{4} \hat\a^a \hat\a_{a}, \nn
(d+ s) \Mh^a&=& \mh\pz\Mh^a-{\w}\Mh^a\pz \mh
+\F\Mh^b\Mh^c+\G\Mh^b\pz\Mh^c.
\ena

By projection in ghost numbers and form degrees, the only nontrivial
terms stemming from these equations give at ghost number 2 the
$s$-transformations of the ghosts as in eq.~(\ref{Q}), and at ghost number 1
the $s$-transformations of the gauge fields as
\bea
s \mz &=& \pzb\cz+\cz\pz\mz-\mz\pz\cz- \shalf \a^a_\z\c_a, \nn
s \Mz^a &=& \mp \pzb m^a +\cz\pz\Mz^a+\w\Mz^a\pz\cz\nn
& &\pm\mz\pz m^a-\w m^a\pz\mz \nn
& &+\F( m^b\Mz^c\pm\Mz^b m^c)+\G( m^b\pz\Mz^c\pm\Mz^b \pz m^c).
\label{stra}
\ena
The rule for the sign in the above formula is that one has an upper
sign when the ghost $m^a$ on the right is commuting (corresponding to an
anticommuting gauge field $\Mz^a$).

It is ensured by construction that the BRST transformation
laws~(\ref{stra}) are nilpotent. Moreover, the projection in the component
formalism of the superfield equations
(\ref{unifN}) turn out to be of the form of these last equations.
The apparent complexity of the latter justifies the $O(N)$ superfield
notation, which captures the whole information about the gauge symmetries
under the form of a single zero curvature condition.

The BRST invariant action is as in eq.~(\ref{act1}).
The gauge fields in the action are the sources of the currents
\EQ
T_{zz}={{\d I_{GF}}\over{\d\mz}}, \qquad
T_{a}={{\d I_{GF}}\over{\d\Mz^a}}.
\EN
These currents would generate the BRST algebra (\ref{Q}) we started from
by a mere application of the OPE technique. They are also the components of
the supercurrent $J_z$ defined in the previous section.

It is useful to illustrate these results explicitly in the cases $N=2,3$
and $4$.

{}For $N=2$, we have the Beltrami differential $\bel$, two gravitini $\a_i,
(i=1,2)$ and one commuting gauge field $\rho$. Their associated ghosts
are denoted as $\cz,\c_i$ and $d$, respectively.
The transformation rules for the ghosts in $N=2$ string are
\bea
s \cz &=& \cz \pz \cz - \frac{1}{4}\c_i \c_i, \nn
s \c_i &=& \cz \pz \c_i - \shalf \c_i \pz \cz - \shalf \e_{ij} \c_j d, \nn
s d &=& \cz \pz d + \shalf \e_{ij} \c_i \pz \c_j,
\label{tran2}
\ena
where $\e_{12}=-\e_{21}=1$ and repeated indices are summed.

{}For $N=3$, the $O(3)$ decomposition gives the Beltrami differential $\bel$,
three gravitini $\a_i$, three gauge fields $\rho_i$, and a fermionic field
$\p$, $(i=1,2,3)$. Their associated ghosts are denoted as $\cz,\c_i, c_i$
and $\d$, respectively.
The transformation rules for the ghosts in $N=3$ string are
\bea
s \cz &=& \cz \pz \cz - \frac{1}{4}\c_i \c_i, \nn
s \c_i &=& \cz \pz \c_i - \shalf \c_i \pz \cz - \shalf \e_{ijk} c_j \c_k, \nn
s c_i &=& \cz \pz c_i + \shalf \e_{ijk} \c_j \pz \c_k - \shalf \c_i \d
 +\frac{1}{4} \e_{ijk} c_j c_k, \nn
s \d &=& \cz \pz \d + \shalf \d \pz \cz - \shalf \c_i \pz c_i,
\label{tran3}
\ena
where $\e_{ijk}$ is the structure constant for $SU(2)$.

{}For $N=4$, the $O(4)$ decomposition gives the Beltrami differential,
four gravitini, six commuting fields that one can assemble into a gauge
field for $O(4)\sim SU(2)\times SU(2)$ rotations, four anticommuting fields
for an internal local supersymmetry and one commuting field
for an internal local $U(1)$ symmetry. Associated with these, we have
the reparametrization ghost $c^z$, four supersymmetry ghosts $\c^a$,
six $SU(2)_{k^+}\times SU(2)_{k^-}$ symmetry ghosts $c_\pm^i$ and one
commuting $\d^a$ and $U(1)$ ghost $d$~\cite{N4,ademollo}.
Here the double index notation $ a=(\a,\bar\a); \a,\bar\a=1,2$ is used
and $i=1,2,3$~\cite{BO}. This corresponds to the so-called large $N=4$ SCA.

The transformation rules for ghosts in $N=4$ string can be written in concise
form as:
\bea
s \cz &=& \cz \pz \cz - \frac{1}{4} \c_a \c^a, \nn
s \c^a &=& \cz \pz \c^a - \shalf \c^a \pz \cz - {{R^{+,i}}_b}^a c_{+,i} \c^b
 - {{R^{-,i}}_b}^a c_{-,i} \c^b, \nn
s c_\pm^i &=& \cz \pz c_\pm^i - \shalf {\e^i}_{jk}c_\pm^j c_\pm^k - \shalf
 (1\mp x) {R^{\pm,i}}_{ab} \pz\c^a \c^b \pm {R^{\pm,i}}_{ab} \d^a \c^b, \nn
s \d^a &=& \cz \pz \d^a + \shalf \d^a \pz \cz
 + \shalf (1+x) {{R^{+,i}}_b}^a \c^b\pz c_{+,i}
 - \shalf (1-x) {{R^{-,i}}_b}^a \c^b\pz c_{-,i} \nn
&& - {{R^{+,i}}_b}^a \d^b c_{+,i} - {{R^{-,i}}_b}^a \d^b c_{-,i}
 - \shalf \c^a\pz d, \nn
s d &=& \cz \pz d - \shalf \d_a \c^a.
\label{tranc}
\ena
The {\it free} parameter $x\equiv \frac{k^+-k^-}{k^++k^-}$ measures the
asymmetry between the two $SU(2)$ current algebras. Its occurrence
in the BRST algebra in component formalism is a
peculiarity of the case $N=4$ and is allowed by the local isomorphism
between $O(4)$ and $SU(2)\times SU(2)$~\cite{N4}.
The $SU(2)$ representation matrices $R^{\pm,i}{}_a{}^b$ have the values
\bea
R^{+,i}{}_{(\a,\bar \a)}{}^{(\b, \bar \b)} =
\left\{ \begin{array}{ll}
\frac{1}{2} \bar \s^i{}_\a{}^\b & \mbox{if $ \bar\a = \bar\b = 1$} \\
\frac{1}{2} \s^i{}_\a{}^\b & \mbox{if $\bar \a =\bar \b =2$} \\
0 & \mbox{otherwise}
\end{array} \right.\ , \ \
 R^{-,i}{}_{(\a,\bar \a)}{}^{(\b, \bar \b)} =
\left\{ \begin{array}{ll}
\frac{1}{2} \bar \s^i{}_{\bar \a}{}^{\bar \b}
 & \mbox{if $ \a = \b = 1$} \\
\frac{1}{2} \s^i{}_{\bar \a}{}^{\bar \b}
 & \mbox{if $\a = \b =2$} \\
0 & \mbox{otherwise}
\end{array}
\right.\ ,
\ena
where $\s^i = (\s^3, \s^+ ,\s^-)$ are the Pauli matrices
in the Cartan basis and $\bar \s^i = ( \s^3, -\s^+,-\s^-)$.
Indices are raised and lowered with the invariant tensors $g^{ij},\eta_{ab}$
and their inverse given by
\bea
g^{+-}=2, \; g^{33}=1; \;\; \eta_{(\a \bar \a) (\b \bar\b)}
= \frac{1}{2} \bar\eta_{\a \b} \bar \eta_{\bar \a \bar \b}\ ,
\ \ {\rm with }\ \ 
\bar \eta_{12} = \bar \eta_{21} =1,\ 
\bar \eta_{11} = \bar \eta_{22} =0.
\ena

The way the classical gauge fields associated to these ghosts transform is
obtained by applying eq.~({\ref{stra}), using the structure
coefficients that one can read from eqs.~({\ref{tran2})-({\ref{tranc}).

It is straightforward to verify on these examples the general results of the
last sections. In particular, the transformation rules for ghosts follow
from eq.~(\ref{sutran}) in terms of $N=2,3$ and $4$ superfields for each
case,\footnote{For $N=4$, the parameter $x$ must be introduced in the ghost
and antighost superfield decomposition.} and the preceding results are
obtained just by projecting on each components.

\sect{The conformal anomaly}

We now give a general formula for the value of the coefficient of conformal
anomaly of the ghost system defined by the action $I_{GF}$. To compute
the anomaly, one can put $M^z_{\z,bg}=0$. Then
\EQ
I_{GF} =\int d^2 z\left(
 b_{zz} \pzb \cz+\sum_{i_1\cdots i_p}\beta_{zi_1\cdots i_p} \pzb
c_\z^{i_1\cdots i_p} \right).
\EN
Since a system of conformal fields $(A,B)$ with Lagrangian
$A\pzb B$ has a conformal anomaly equal to $ \pm 2(6n^2-6n+1)$ where $n$ is
the conformal weight of the field $A$ and the sign $+$ ($-$) occurs if
$A$ and $B$ commute (anticommute)~\cite{FMS}, the value of the conformal
anomaly associated to the action $I_{GF}$ is
\EQ
c(N)=\ \sum_{p=0}^N\ (-)^{p+1}{}_NC_p \ 2\left(6w(p)^2-6w(p)+1\right),
\EN
where $w(p)={p\over2}-1$.

One has $c(0)=-26, c(1)=-15, c(2)=-6$, which express the well-known
fact that superstrings for $N=0$, $1$ and $2$ have critical dimensions
$D=26$, $10$ and $4$ respectively, and, for all values of $N\geq 3$,
\EQ
\label{an}
c(N)=0,\qquad N \geq 3.
\EN
This formula can be proved algebraically by using the defining relation
$(1+z)^N=\sum_{p=0}^N{}_NC_p z^p$. In the next section we will demonstrate
it in another way which emphasizes the string embedding property.
Eq.~(\ref{an}) indicates that all superstrings based on the full $O(N)$
superspace have vanishing critical dimension for $N\geq 3$~\cite{schou,BO,BOP},
and can therefore only be of a purely topological nature.

Let us also consider the question of finding the anomaly, defined as the
local $3$-form solution modulo $\hat d$-exact terms of the consistency equation
$\int \hat d \hat \Delta_3=0$. The conventional consistent anomaly,
which is the possible local counterterm which can possibly break the
conformal BRST Ward identity, is the component $\Delta^1_2$ with ghost number
1 of $\hat\Delta_3$, defined modulo $d$- and $s$-exact terms.

In the case of $N=0$ supersymmetry, one has
\EQ
N=0\ :\ \hat \Delta_3=\hat \mu^z\pz \hat \mu^z\pz^2\hat \mu^z.
\EN
This expression is in correspondence with the property that the violation
of the conservation of the holomorphic component of the energy-momentum tensor
is  proportional to $\pz^3 \mz$ \cite{BB}.

To generalize this expression to higher values of rank $N$, we must search
for the possible completion of $\hat \mu^z\pz \hat \mu^z\pz^2\hat \mu^z$ by
terms such that the whole expression is $\hat d$-closed but not
$\hat d$-exact. This can be achieved by finding zero-forms $\Delta^3_0$
with ghost number three which must be BRST-closed, and contain
$c^z\pz c^z \pz^2 c^z$.

Quite interestingly, power counting requirements ($\Delta_0^3$ must be made
of local terms with the same canonical dimension as $c^z\pz c^z\pz^2 c^z$)
imply that we only have {\it local} solutions, i.e, solutions
which involve only positive powers of the derivative $\pz$ for $N=0,1,2,3$.
Indeed, for higher values of $N$, power counting forbids $\Delta_0^3$ to
depend on the ghosts whose weight is bigger or equal to one, and thus
the consistency equation cannot be satisfied.

We find the following expressions for the consistent anomalies in components:
\bea
N=1&:&\hat \Delta_3 = \hat \mu^z\pz \hat \mu^z\pz^2\hat \mu^z - \hat \mu^z
 (\pz \hat \a^z)^2 + \shalf \pz \hat\mu^z \hat \a^z \pz \hat \a^z, \nn
N=2&:& \hat \Delta_3 = \hat \mu^z\pz \hat \mu^z\pz^2\hat \mu^z - \hat \mu^z
 (\pz \hat \a^z_i)^2 + \shalf \pz \hat\mu^z \hat \a^z_i \pz \hat \a^z_i
 + \hat \mu^z \hat \rho^z \pz \hat \rho^z - \shalf \e_{ij} \hat \rho^z
 \hat \a^z_i \pz \hat \a^z_j, \nn
&& \hspace{10mm} (i,j=1,2),\nn
N=3&:&\hat \Delta_3 = \hat \mu^z\pz \hat \mu^z\pz^2\hat \mu^z - \hat \mu^z
 (\pz \hat \a^z_i)^2 + \shalf \pz \hat \mu^z \hat \a^z_i \pz \hat \a^z_i
 + \hat \mu^z \hat \rho^z_i \pz \hat \rho^z_i - \hat \mu^z
 (\hat \p^z)^2 \nn
&& - \shalf \hat \rho_1^z \hat \rho^z_2 \hat \rho^z_3
 - \shalf \e_{ijk} \hat \rho^z_i \hat \a^z_j \pz \hat \a^z_k
 + \shalf \hat \rho_i^z \hat \a^z_i \hat \p^z ,\;\;
 (i,j,k=1,2,3),
\ena
where $\hat\a,\hat\rho,\hat\p$ are the unified fields as defined in
eq.~(\ref{unified}) for gravitini, gauge and fermionic fields.

The N=4 case is subtle because of the existence of a generator of dimension
zero. For all other values of $N>4$,
there is no consistent anomaly. This is compatible with the fact that the
sums of all ghost anomaly contributions vanish as shown in our computation
in the first part of this section and also with the absence of
central extension for $N >4$~\cite{schou}.
This property probably expresses that the conformal field theory is
free of infinities for $N>4$.

\sect{The superstring embedding}

The ghost system of the $O(N)$ superspace is made of a tower of ghosts
which describe the antisymmetric tensors of $O(N)$, with $_NC_p$ ghosts
with weight $w(p)$ and statistics $(-1)^{p+1}$.
We will show the interconnection of these systems when $N$ varies.

Let us define the following notation which defines the $N$-superstring from
the knowledge of its ghost spectrum:
\EQ
N{\rm -theory}\ \equiv\ \{\ (_NC_p,w(p))\ ;\ 0\leq p \leq N\ \},
\EN
where the first entry indicates degeneracy and the second the
conformal weight. The antighosts are implicitly defined as the duals of the
ghosts, and the action is as in (\ref{act1}).
The theory of rank $N-1$ is thus represented by
\EQ
(N-1){\rm -theory}\ \equiv\ \{\ (_{N-1}C_p,w(p))\ ;\ 0\leq p \leq N-1\ \}.
\EN

Thanks to this notation, it becomes almost obvious to see that embedding of
the $(N-1)$-theory into the $N$-theory just follows from the relation
\EQ
\label{com}
_NC_p=\ _{N-1} \hspace{-.2mm} C_p+\ _{N-1} \hspace{-.2mm} C_{p-1},
\EN
valid for all values of $N\geq 1$ and $p\geq 1$.

Indeed, this relation suggests considering the ghost system obtained by
isolating in the $N$-theory one ghost with weight $w(1)$, $_{N-1}C_1$
ghosts with weight $w(2)$, $\cdots$, $_{N-1}C_{p-1}$ ghosts with weight
$w(p)$, and so on down to $_{N-1}C_{N-1}=1$ ghost with weight $w(N)$.
We can denote this subsystem as
\EQ
\Delta N{\rm -theory}\ \equiv\ \{\ (_{N-1}C_{p-1},w(p))\ ;\ 1\leq p
\leq N-1\ \}.
\EN
But then, one has obviously from eq.~(\ref{com})
\bea
N{\rm -theory}\ \equiv\ (N-1){\rm -theory}\ \cup \ \Delta N-{\rm theory}.
\ena

For $N=1,2,3$, the $ \Delta N$-theory are anomalous.
More precisely: The $\Delta 1$-theory is made from one $(\b,\c)$
ghost-antighost pair and has $c=11$; the $\Delta 2$-theory is made from one
$(\b,\c)$ pair and one Grassmann-odd ghost-antighost pair contributing to the
anomaly by the amount $-2$; this theory has thus $c=11-2=9$; finally, the
$\Delta 3$-theory is made from one $(\b,\c)$ pair, two odd pairs
contributing to the anomaly by the amount $c=-2\times 2$ and one
even pair contributing to the anomaly by the amount $c=-1$;
the $\Delta 3$-theory has thus $c=11-4-1=6$.

For $N\geq 4$, one encounters a new regime. The $\Delta 4$-theory is made
from one $(\b,\c)$ pair, three odd pairs contributing to the anomaly by
the amount $c=-2\times 3$, three even pairs contributing to the anomaly
by the amount $c=-1\times 3$ and one odd pair contributing to the anomaly
by the amount $c=-2$; the $\Delta 4$-theory has thus $c=11-6-3-2=0$.
Since, on the other hand, we can easily verify that
for the $3$-theory $c=0$, we see that the $4$-theory is also anomaly free,
which could of course be verified directly. Moreover, when computing
the path integral over the ghost and antighost fields,
since both the $\Delta 4$-theory and $3$-theory are separately
anomaly free, the fields of the $\Delta 4$-theory can be integrated out,
and we obtain the result that the $3$-theory is embedded in the $4$-theory.

This can be pursued by induction. One can prove in this way
(i) that the $N$-theory as well as the $\Delta (N+1)$-theory are
anomaly free for $N\geq 3$ and
(ii), as a corollary, that the $N$-theory is embedded in the $(N+1)$-theory.

\sect{Connection to purely topological actions}

Let us conclude by noting that the $(N,N')$ theories can be also viewed as
topological 2D-gravity coupled to topological sigma models with bosonic and
fermionic coordinates.
The field spectrum of the left sector of a worldsheet with $(N,N')$ local
supersymmetry is made of $2^N$ classical gauge fields (including the Beltrami
differential) that we denoted as $C^{1-p/2}_{i_1\ldots i_p}$, $1\leq p\leq N$,
and of $2^N$ ghosts, with an equal partition between bosons and fermions.
One has a similar situation in the right sector, by replacing $N$ by $N'$.
For $N=1$, it has been shown in refs.~\cite{BGR,BA} that the $(1,1)$
2D-supergravity BRST algebra can be twisted into that of pure topological
2D-gravity by field redefinitions mixing the gauge fields and the ghosts,
and concluded that the $N=1$ superstring can be viewed as topological
2D-gravity coupled to a topological sigma model. Quite remarkably this
observation can be extended to the case of any given $(N,N')$ local
supersymmetry. Indeed, in the left sector, one can do field redefinitions
involving the $2^N$ classical gauge fields and their ghosts to obtain
$2^{N-1}$ sets of topological pairs $Y^i$, $F^i$, $1\leq i \leq 2^{N-1}$,
such that $s Y^i = F^i$, $s F^i =0$. For $N\geq 2$, the $2^{N-1}$ fields
$Y^i$ consists of the Beltrami differential, $2^{N-2}-1$ bosons and $2^{N-2}$
fermions which can thus  be interpreted as the coordinates of a topological
sigma model. (For $N=1$, only the Beltrami differential and its
topological ghost occur.)

These field redefinitions, which modify the ghost numbers as well as the
conformal weights, are such that the bosonic and fermionic ghosts $F^i$
are quadratic products of some of the fields $C^{1-p/2}_{i_1\ldots i_p}$ by
some of the ghosts $c^{1-p/2}_{i_1\ldots i_q}$. They can be read off from
the general formula for the BRST transformation of $(N,N')$ local
supersymmetry. This is a mere generalization of getting the topological ghost
$\Psi^z_\z$ in topological 2D-gravity as the product of the gravitino
$\a^\shalf_\z$ by its commuting ghost $\c^\shalf$,
$\Psi^z_\z=\a^\shalf_\z \c^\shalf$.

One can complete this picture by finding a ghost of ghost phenomenon which
also generalizes that of the case $N=1$, where the ghost of ghost for
reparametrization is $\Phi^z= \c^\shalf\c^\shalf$. This ghost of ghost
phenomenon simply takes into account all internal symmetries of
$O(N)$ superspace.

It is then possible to redefine the antighosts as in refs.~\cite{BGR,BA},
in a way which is the conjugate to the redefinitions of the ghosts.
This amounts altogether to canonical transformations. In this way, we
finally arrive at the conclusion that the BRST invariant action for the pure
worldsheet theories with $(N,N')$ local supersymmetry can be also considered
as topological 2D-gravity coupled to a topological sigma model with
$2^{N-2}-1$ bosonic coordinates and $2^{N-2}$ fermionic coordinates in the
left sector, and $2^{N'-2}-1$ bosonic coordinates and $2^{N'-2}$ fermionic
coordinates in the right sector.

\vs{5}
\noindent
{\bf Acknowledgement:}
Some of the calculations in this paper were done by using the OPE package
developed by K. Thielemans, whose software is gratefully acknowledged.

\newcommand{\NP}[1]{Nucl.\ Phys.\ {\bf #1}}
\newcommand{\PL}[1]{Phys.\ Lett.\ {\bf #1}}
\newcommand{\CMP}[1]{Comm.\ Math.\ Phys.\ {\bf #1}}
\newcommand{\PR}[1]{Phys.\ Rev.\ {\bf #1}}
\newcommand{\PRL}[1]{Phys.\ Rev.\ Lett.\ {\bf #1}}
\newcommand{\PTP}[1]{Prog.\ Theor.\ Phys.\ {\bf #1}}
\newcommand{\PTPS}[1]{Prog.\ Theor.\ Phys.\ Suppl.\ {\bf #1}}
\newcommand{\MPL}[1]{Mod.\ Phys.\ Lett.\ {\bf #1}}
\newcommand{\IJMP}[1]{Int.\ Jour.\ Mod.\ Phys.\ {\bf #1}}
\newcommand{\JP}[1]{Jour.\ Phys.\ {\bf #1}}
\newcommand{\JMP}[1]{Jour.\ Math.\ Phys.\ {\bf #1}}

\end{document}